\documentclass[pre,aps,twocolumn, superscriptaddress]{revtex4}
\usepackage{amsmath}
\usepackage{epsfig}
\usepackage{amssymb}
\usepackage{color}

\newcommand {\La}{${\text L}_\alpha^C$} 
\newcommand {\Hi}{${\text H}_{II}^C$}
\newcommand {\abs}[1]{\left |#1 \right | }
\newcommand{\subfig}[1]{({\bf \uppercase{#1}})}

\begin{document}

\title{The Thermodynamics of Endosomal Escape and DNA Release from
Lipoplexes}

\author{Yotam Y. Avital}
\affiliation{Department of Biomedical Engineering, Ben-Gurion University of the
Negev, Be'er Sheva 85105, Israel}
\affiliation{Ilae Katz Institute for Nanoscale Science and Technology, Ben-Gurion University of the
Negev, Be'er Sheva 85105, Israel}

\author{Niels Gr\o nbech-Jensen}
\affiliation{Department of Mechanical and Aerospace Engineering, University of California, Davis,
California 95616, USA}
\affiliation{Department Mathematics, University of California, Davis,
California 95616, USA}

\author{Oded Farago}
\affiliation{Department of Biomedical Engineering, Ben-Gurion University of the
Negev, Be'er Sheva 85105, Israel}
\affiliation{Ilae Katz Institute for Nanoscale Science and Technology, Ben-Gurion University of the
Negev, Be'er Sheva 85105, Israel}

\begin{abstract}

Complexes of cationic and neutral lipids and DNA (lipoplexes) are
emerging as promising vectors for gene therapy applications. Their
appeal stems from their non pathogenic nature and the fact that they
self-assemble under conditions of thermal equilibrium. Lipoplex
adhesion to the cell plasma membrane initiates a three-stage process
termed transfection, consisting of (i) endocytosis, (ii) lipoplex
breakdown, and (iii) DNA release followed by gene expression. As
successful transfection requires lipoplex degradation, it tends to be
hindered by the lipoplex thermodynamic stability; nevertheless, it is
known that the transfection process may proceed spontaneously. Here,
we use a simple model to study the thermodynamic driving forces
governing transfection. We demonstrate that after endocytosis [stage
(i)], the lipoplex becomes inherently unstable.  This instability,
which is triggered by interactions between the cationic lipids of the
lipoplex and the anionic lipids of the enveloping plasma membrane, is
entropically controlled involving both remixing of the lipids and
counterions release. Our detailed calculation shows that the free
energy gain during stage (ii) is approximately linear in $\Phi_+$, the
mole fraction of cationic lipids in the lipoplex. This free energy
gain, $\Delta F$, reduces the barrier for fusion between the
enveloping and the lipoplex bilayers, which produces a hole allowing
for DNA release [stage (iii)]. The linear relationship between $\Delta
F$ and the fraction of cationic lipids explains the experimentally observed
exponential increase of transfection efficiency with $\Phi_+$ in
lamellar lipoplexes.

\end{abstract} 

\maketitle

Somatic gene therapy holds great promise for future medical
applications including, for example, new treatments for various
inherited diseases and cancers \cite{Tampelton:2000}. Within this
approach, an attempt is made to replace damaged genes with properly
functioning ones. The core of the process, called transfection,
includes the key steps of transferring foreign DNA into a target cell,
followed by expression of the genetic information. Complexes composed
of cationic lipids (CLs) and DNA, designated lipoplexes, constitute
one of the most promising non-viral gene delivery systems
\cite{Felgner:1996,Miller:1998,Mahato:2002}. Though their transfection
efficiency (TE) is, in general, inferior to that of viral vectors,
lipoplexes have the advantage of triggering low immune response, and
being non-pathogenic
\cite{Mahato:2002,Huang:2005,Ewert:2005,Kamiya:2001}. Furthermore,
lipoplexes allow transfer of larger DNA segments. Their production
does not require sophisticated engineering, since they form
spontaneously in aqueous solutions when DNA molecules are mixed with
CLs and neutral lipids (NLs)
\cite{Radler:1997,Koltover:1998,gjf-jacs,gjf-soft}. The main
thermodynamic driving force for lipoplex formation is the entropic
gain stemming from the release of the tightly bound counterions from
the DNA and the lipid bilayers. X-ray diffraction experiments have
revealed several liquid crystalline phases of CL-DNA complexes. The
two most prominent structures are: (i) a lamellar phase (\La), with 2D
smectic array of DNA within lipid bilayers \cite{Radler:1997}, and
(ii) an inverted hexagonal phase (\Hi), where the DNA rods are packed
in hexagonal lattice and the lipids form monolayers around them
\cite{Koltover:1998}.

Isoelectric complexes, where the total charge on the DNA molecules
exactly matches the total charge of the CLs, are the most stable ones
because they enable nearly complete counterion release
\cite{Harries:1998}. The thermodynamic stability of a lipoplex,
however, is only one of several biophysical parameters affecting the
TEs of lipoplexes.  Another parameter is the liquid crystalline
structure of the complex \cite{Ewert:2004}, which is largely
determined by the bending rigidity and spontaneous curvature of the
lipids \cite{Safinya:2001}. Generally speaking, \Hi~complexes exhibit
higher TEs than \La~complexes. A third parameter is the charge density
(per unit area) of the lipoplex membranes, which can be varied by
mixing different ratios of CLs and NLs, and by using multivalent CLs
\cite{Ahmad:2005}.

Further improvement of the therapeutic efficacy of lipid vectors
requires better understanding of their mechanism of transfection, and
the biophysical parameters of the CL-DNA complexes that influence
it. Transfection is viewed as a three-stage process starting with
adsorption and entry (via endocytosis) of the CL-DNA complex into the
cell, followed by lipoplex degradation, and finally ending with the
release of the DNA, making the latter available for expression
\cite{Ewert:2004,Huebner:1999,Kennedy:2000}. The first stage is driven
by electrostatic attraction between the lipoplex CLs and the
negatively charged lipids of the cell plasma membrane, which enables
further release of counterions (see discussion below). After
endocytosis the complex is within the cell, trapped inside an
endosome. The second stage of the transfection process, which often
emerges as the rate-limiting one, involves breakdown of CL-DNA
complex. During this stage, the endosomal and the lipoplex external
membranes fuse \cite{Ewert:2004}. It has been speculated that the
improved TE of hexagonal complexes over lamellar ones stems from its
lower energy barrier of fusion \cite{Ewert:2004}.  In the case of
lamellar complexes, the fusion energy barrier decreases (and TE
increases) when the mole fraction of CLs increases. These observations
suggest that the electrostatic attraction between the lipoplex and the
endosomal membrane triggers a thermodynamic instability leading to
morphologic changes. In this work we explore the thermodynamic driving
forces governing the transfection process from the stage of adhesion
and endocytosis, up to the stage of DNA release.

\begin{figure*}
   \centering\includegraphics[width=0.9\textwidth]{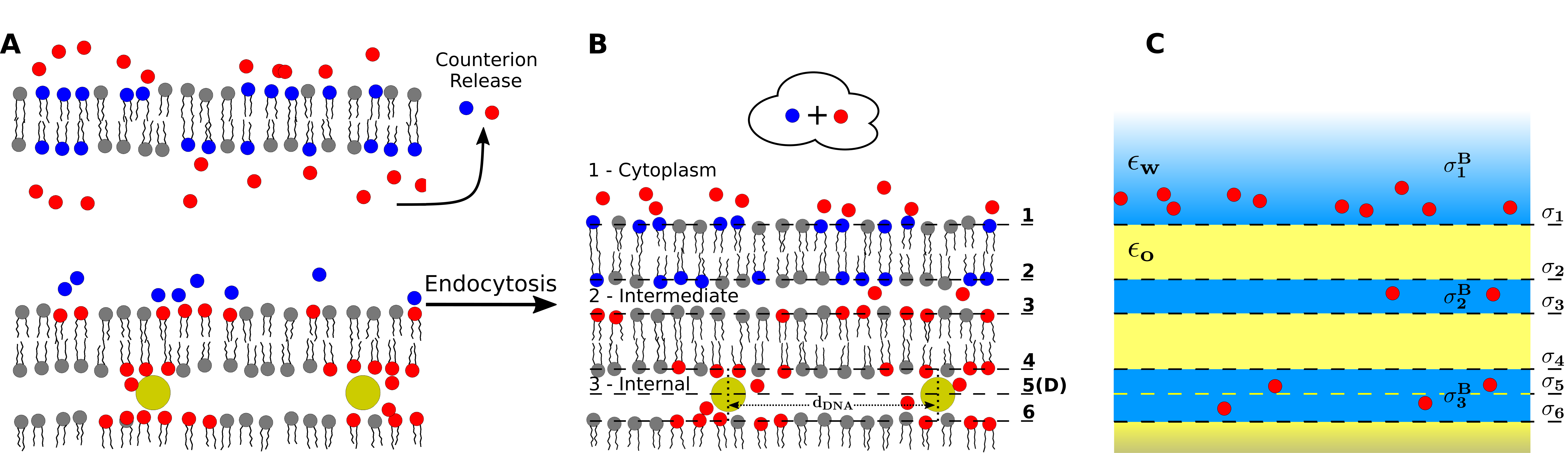} 
    \caption{{\bf (A)} Schematics of a complex of CLs (head groups
    depicted as red circles), NLs (head groups - grey circles), and
    DNA rods (larger yellow circles), separated from the plasma
    membrane which is composed of ALs (head groups - blue circles) and
    NLs. The lipoplex attracts a layer of bound anions (shown as blue
    circles), while the plasma membrane is surrounded by bound cations
    (red circles). {\bf (B)} The state of the system after adhesion
    and endocytosis, the formation of which is driven by cation-anion
    pairs release.  {\bf (C)} A simplified model of the system
    depicted in {\bf B} (see detailed explanation in text). The model
    system consists of 6 uniformly charged plates with charge density
    $\sigma_i$ and three water layers (shown in blue) where the ions
    reside. The yellow stripes represent hydrophobic regions that do
    not include ions, and at which the electric filed must
    vanish. Notice that the 5th charged plate, which represents the
    DNA array, allows crossover of ions.}
    \label{fig:fig1}
\end{figure*}

CL-DNA complexes adhere to cell membranes due to considerations
similar to those triggering their formation, namely counterion
release. Both the plasma membrane and the external bilayer of the
lipoplex are covered with layers of tightly bound counterions
[fig.~\ref{fig:fig1}\subfig{a}]. These counterions neutralize the
lipid charges, and exclude the electric field from the oily parts of
the membranes. The loss of positional entropy of the bound counterions
is significantly lower than the energetic cost of allowing an electric
field to penetrate the bilayers low dielectric hydrophobic core. When
the oppositely charged surfaces are in close proximity, the anionic
and cationic lipids can neutralize each other, which enables the
release of counterion pairs. The positional entropy gained by the
released counterions is the main driving force for cell-lipoplex
adhesion which initiates cellular entry via endocytosis.

Figure 1\subfig{b} shows, schematically, a small segment of a lipoplex
trapped within an endosome. The entrapped lipoplex represents a
thermodynamic system that is substantially different from the lipoplex
originally residing outside the cell.  The difference stems from the
presence of anionic lipids (ALs) in the plasma membrane which can now
mix with the CLs and NLs of the lipoplex \cite{Ewert:2004}. The
process of lipid mixing is slow since it requires the lipids to
``flip-flop'' between monolayers; nevertheless, it encompasses a large
entropic reward. Moreover, redistribution of the lipids, while
protecting the hydrophobic cores of the bilayers from electric fields,
dictates that the counterions ``escort'' the flip-flopping charged
lipids. When ions move between the different aqueous layers
of the system they meet oppositely charged ions which allows
them to mutually leave the system without affecting its charge
neutrality.

The considerations presented in the previous paragraph suggest that
the entrapment of a lipoplex by the endosome may be sufficient to
render it thermodynamically unstable. This is obviously a desirable
feature since the ultimate goal of the transfection process is
lipoplex disassembly and DNA release. To better understand the
thermodynamics of transfection, we will present a simplified model
where electrostatic interactions are considered within the framework
of a mean field approximation. The model treats the membranes, as well
as the DNA array, as uniformly charged planner sheets
[fig.~\ref{fig:fig1}\subfig{c}].

Before describing the model, let us return to figure
\ref{fig:fig1}\subfig{b} illustrating the entrapped lipoplex
immediately after endocytosis.  The system constitutes six charged
layers. In reverse order [from number 6 to 1, see
fig.~\ref{fig:fig1}\subfig{b}], these charged layers correspond to: 6
- the lipid monolayer ``below'' the DNA array, 5 (also denoted by D) -
the DNA array, 4 - the lipid monolayer ``above'' the DNA array, 3 and
2 - the ``intermediate'' lipid monolayers, and 1 - the lipid monolayer
facing the cytoplasm. The three aqueous environments in the system
will be denoted by: 1 - the cytoplasm, 2 - the intermediate thin water
layer between the endosomal membrane and the lipoplex, and 3 - the
internal water region surrounding the first DNA layer. At the initial state,
the lipid composition in monolayers 1 and 2 is that of the cell plasma
membrane. It consists of ALs and NLs only and, for simplicity, will be
assumed to be symmetric. Similarly, surfaces 3-6 are in the
equilibrium state of the self-assembled lipoplex, and have the same
CL to NL ratio. We will also assume that the NLs of the plasma and
lipoplex membranes are of the same type.

The three major contributions to the free energy of the system arise
from electrostatic interactions, lipid mixing entropy, and the entropy
loss of bound counterions. In our model system,
fig.~\ref{fig:fig1}\subfig{c}, the lipid monolayers are replaced with
uniformly charged flat surfaces of charge density $\sigma_i$
($i$=1,2,3,4,6). The aqueous solutions have a dielectric constant
$\epsilon_w\simeq80$, while that of the hydrophobic regions,
$\epsilon_o$ is assumed to be vanishingly small. This precludes the
penetration of electric fields into the hydrophobic regions due to the
associated very large electrostatic energy
\cite{Kiometzis:1989,Winterhalter:1992}. (We note that the cytoplasam
is occupied with concentrated macromolecules. Their presence changes
the inside relative permittivity to values ranging from about 50 to
over 200 \cite{davey:1995,gimsa:1996}, for which the assumption
concerning the exclusion of the electric field is from the hydrophobic
regions still holds.) A somewhat greater approximation that we make is
replacing the electric field of the DNA array with the electric field
of a flat surface of charge density per unit area $\sigma_5 =
\lambda/{d_{\rm DNA}}$ where $\lambda \simeq 1.7e/{\rm \AA}$ is the
linear (per unit length) charge density of the DNA rod, and $d_{\rm
DNA}$ is the inter-DNA spacing. A more detailed mean field
calculation, taking into account the geometry of the DNA rods, can be
performed computationally \cite{Harries:1998}.  Such a calculation,
however, is not necessary here. In order to understand the ``big
picture'', one only needs to recognize that the counterions must
arrange themselves to minimize the electrostatic energy. Any
appreciable deviation in the ions distribution will involve an energy
cost much larger than the entropic components of the free
energy. Specifically for the model system in
fig.~\ref{fig:fig1}\subfig{c}, the number of ions per unit area
present in each aqueous environment will have to match the area charge
densities of the surfaces in a manner that eliminates the electric
field from the low dielectric regions \cite{footnote1}. Electric field
can be present in the aqueous regions, and the associated energy can
be derived by integrating over the electrostatic energy density. Under
no-salt conditions, this precisely gives the free energy cost
attributed to the bound counterions.  We will not perform the exact
calculation (which requires the solution of a corresponding
Poisson-Boltzmann equation), but instead employ a simple approximation
and assign each bound counterion with a free energy of $1k_BT$
\cite{Harries:1998,Harries:2000}.

Let us denote by $\phi_i^+$, and $\phi_i^-$ the mole fractions of the
cationic and anionic lipids in monolayer $i$, respectively, where the
monolayers are located at $z_i$. The area per lipid, $a$, is taken as
identical for all three lipid species (CLs, ALs, and NLs). We denote
by $n^+$ and $n^-$ the number densities, per unit volume, of the
cations and anions, respectively.  To make the mean field
approximation applicable, we consider the case where all the charged
lipids, as well as the counterions, are monovalent. Assuming ideal
lipid mixing in the monolayers, the uniform charge density of each
surface is $\sigma_i = e \left ( \phi_i^+ - \phi_i^- \right)/a$ where
$e$ is the electron charge.  Since the system has a planar symmetry in
the $x-y$ plane, the electric field at any point must be orthogonal to
the plane, i.e., along the $z$ axis. Moreover, $n^+=n^+(z)$,
$n^-=n^-(z)$, and both vanish inside the hydrophobic parts of the
membranes [colored in yellow in fig.~\ref{fig:fig1}\subfig{c}] where
$\epsilon_o \ll \epsilon_w$. The electric field at a given coordinate
$z$ is given by $E_z=\tilde {\sigma} /2 \epsilon_z$ where
\begin{eqnarray} 
    &\tilde {\sigma}& = \nonumber \\  
    && e \int_{\infty}^z \left [ n^+ \left (z' \right ) - n^- \left (z' \right )
        +\sum_{i=1}^{6} \sigma_i \delta \left ( z' - z_i \right) \right] dz' \nonumber
    \\ 
    &-& e \int_z^{-\infty} \left [ n^+ \left (z' \right ) - n^- \left (z' \right
        ) +\sum_{i=1}^{6} \sigma_i \delta \left ( z' - z_i \right) \right] dz' =
    \nonumber \\ 
    &=& 2 e \int_{\infty}^z \left [ n^+ \left (z' \right ) - n^- \left (z' \right
        ) +\sum_{i=1}^{6} \sigma_i \delta \left ( z' - z_i \right) \right] dz', 
    \nonumber \\ 
    \label{eq:sigma} 
\end{eqnarray} 
and $\epsilon_z$ is the dielectric constant at $z$. The second equality in
eq.~\eqref{eq:sigma} is due to the overall charge neutrality of the system. 

The requirement that the electric field vanishes inside the low
dielectric regions of the bilayers can be used to determine, via
eq.~\eqref{eq:sigma}, the number of bound counterions, $N_j^B$, in the
three aqueous solutions of the system ($j=1,2,3$). In each such region
we expect to find only one type of counterions since pairs of
oppositely charged counterions can be released without affecting the
charge balance. Defining $\sigma_j^B = s_j e\left(N_j^B /a\right)$,
where $s_j=+1$ ($s_j=-1$) for cations (anions), the number of
counterions bound to the endosome on its cytoplasmic side is
calculated through
\begin{equation}
    \sigma_1^B = -\sigma_1.
    \label{eq:sigma_b1}
\end{equation}
This relation ensures that the electric field between layers $i=1$ and $i=2$
vanishes. By the same logic, in the intermediate water layer 
\begin{equation}
    \sigma_2^B = - \left (\sigma_1 + \sigma_2 + \sigma_3 + \sigma_1^B \right )
    =- \left ( \sigma_2 + \sigma_3 \right),
    \label{eq:sigma_b2}
\end{equation}
and in the internal water layer
\begin{equation}
    \sigma_3^B =  - \left ( \sigma_4 + \sigma_5  + \sigma_6 \right).
    \label{eq:sigma_b3}
\end{equation}

At short times after cellular intake, the surface charge densities of
the endosome layers, $\sigma_i $, match those of the cell plasma
membrane $(i=1,2)$, and the lipoplex $(i=3-6)$. This initial state is,
however, no longer the equilibrium state, since the anionic and
cationic lipids can now mix with each other. This occurs via slow, but
steady, ``flip-floping'' events switching lipids between monolayers
$i=1-4$ \cite{footnote2}. The redistribution of lipids between these
monolayers not only increases the mixing entropy of the lipids within
the layers, but may also allow further release of counterions whose
densities within the aqueous solutions are simultaneously updated in
order to satisfy the conditions of
eqs.~\eqref{eq:sigma_b1}-\eqref{eq:sigma_b3}. Taking these
considerations into account, we write the total free energy of the
system, per unit area of the lipids $a$, as
\begin{widetext}
\begin{equation}
    \frac{F}{a k_B T}  = \sum_{i=1}^4 \left [ \phi_i^+ \log \left (\phi_i^+
        \right)  + \phi_i^- \log \left ( \phi_i^- \right )  + \left ( 1-
            \phi_i^+ - \phi_i^- \right ) \log \left (1-\phi_i^+ - \phi_i^-
        \right ) \right ] + \sum_{j=1}^3 N_j^B, \label{eq:free_energy}
\end{equation}
\end{widetext}
where $\phi_i^{\pm}$ are the mole fractions of cationic ($+$) and
anionic ($-$) lipids at the $i$-th layer, and $N_j^B$ is the number of
bound counterions per unit area $a$ at the $j$-th water layer (see
definitions also above). The first term in eq.~\eqref{eq:free_energy}
accounts for the mixing entropy of the lipids in each monolayer, while
the second term represents the entropy cost of bound counterions. The
former is based on the mean field assumption of ideal mixing.  The
latter employs the commonly used assumption of $1 k_B T$ per bound
counterion.

Let $\{\phi_{i,0}^\pm\}$ denote the initial mole fractions of the CLs
and ALs. To find the equilibrium state, we need to minimize the free
energy in eq.~\eqref{eq:free_energy} with respect to the variables
$\{\phi_i^\pm\}$, under the constraints that $\sum_{i=1}^4 \phi_i^+ =
\sum_{i=1}^4 \phi_{i,0}^+$, and $\sum_{i=1}^4 \phi_i^- = \sum_{i=1}^4
\phi_{i,0}^-$ representing the preservation of the total number of
lipids of each type. The dependence of $\{N_j^B\}$ on the variables
$\{\phi_i^\pm\}$, is given by
eqs.~\eqref{eq:sigma_b1}-\eqref{eq:sigma_b3}, where $N_j^B = (a
/e)\abs{\sigma_j^B}$, and $\sigma_i = e \left ( \phi_i^+ - \phi_i^-
\right)/a$. Notice that in contrast to the lipids, the total number of
bound counterions is not fixed but may vary by intake or release of
ions from the cytoplasm.

\begin{figure}
    \centering
    \includegraphics[width=0.4\textwidth]{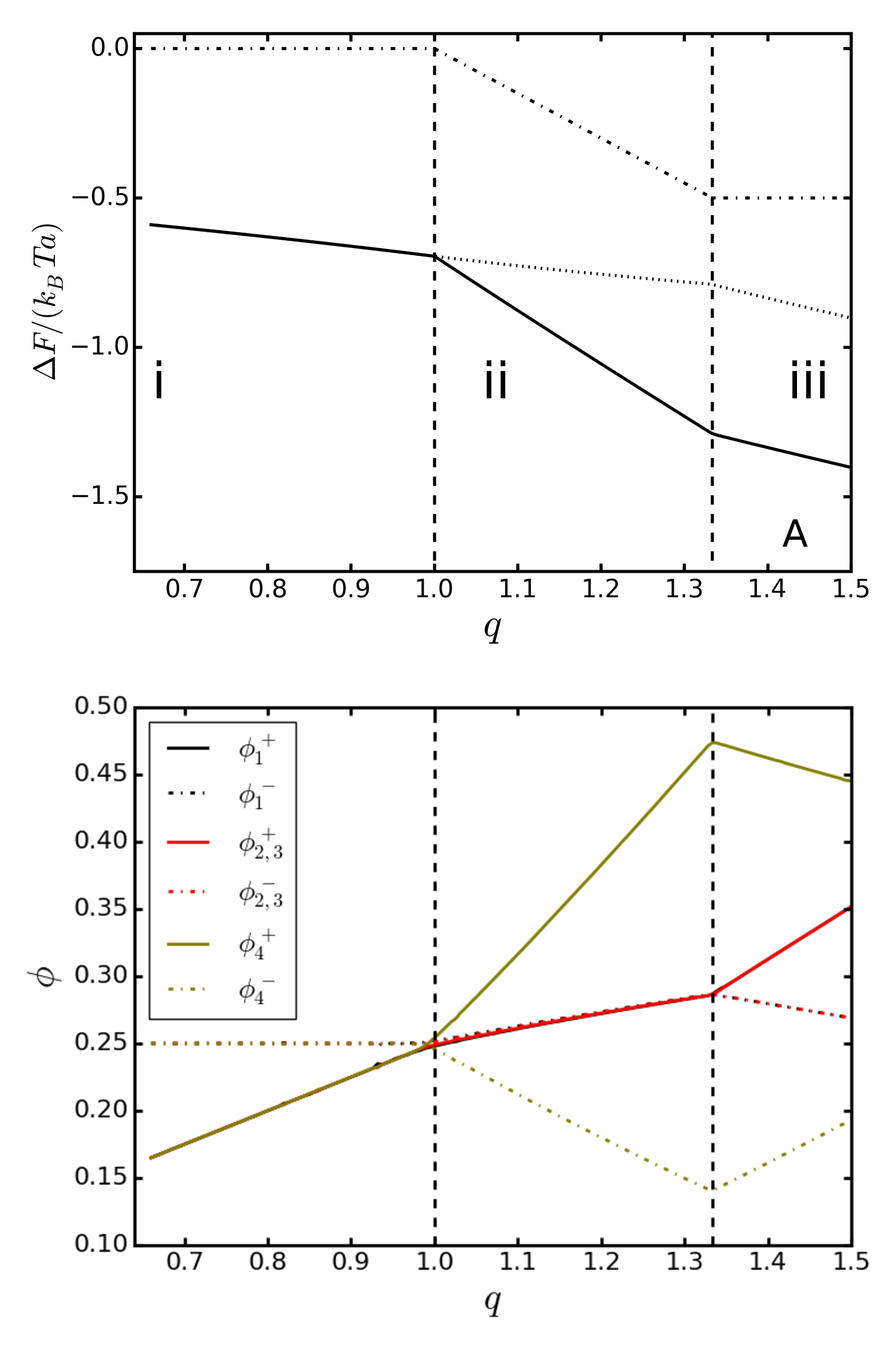}
    \caption{{\bf (A)}. Solid line - the free energy $\Delta F$ (see
    text for definition), as a function of $q=2\Phi_+/\Phi_D$ with
    $\Phi_D=1$ and $\Phi_- = 0.5$.  Dot-dashed and dotted lines show
    the partial contributions to $\Delta F$ originating, respectively,
    from counterion release and lipid mixing.  {\bf (B)}. The
    equilibrium distribution of CLs (solid lines) and ALs (dot-dashed
    lines) in monolayers $i=1$ (black), $i=2,3$ (red) and $i=4$
    (yellow).  The vertical dashed lines mark the transition points
    between the different regimes discussed in the text.}  
    \label{fig:fig2_iso_entropy}
\end{figure}

The free energy $\Delta F$, per unit area $a$, that the system may
gain during stage (ii) of the transfection process is given by the
difference in $F$ [eq.~\eqref{eq:free_energy}] between the equilibrium
and initial states. In the initial state, the distribution of lipids
in the plasma membrane is given by $\phi_{i,0}^- =\Phi_{-}$ and
$\phi_{i,0}^+ = 0$, for $i=1,2$. In the lipoplex membranes
($i=3,4,6$), $\phi_{i,0}^- = 0$ and $\phi_{i,0}^+=\Phi_+$. For
convenience, we define the ``mole fraction'', $\Phi_D = -(a/e)
\sigma_5$, associated with the DNA array. Figure
\ref{fig:fig2_iso_entropy}\subfig{a} plots our results for $\Delta F$
as a function of $q = 2 \Phi_+ / \Phi_D$, which is the lipoplex charge
ratio. The ratio $q$ is varied by changing $\Phi_+$, while keeping
$\Phi_D=1$ fixed. We also fix $\Phi_{-}$, the initial anionic lipid
mole fraction in the plasma membrane, to 0.5. The data for $\Delta F$
is plotted in solid line, while the dotted and dashed curves show,
respectively, the partial contributions due to lipid mixing [first
term in eq.~\eqref{eq:free_energy}] and the bound counterions (second
term). The results revel the existence of three different regimes. In
regime (i), corresponding to $q<1$, the decrease in $\Delta F$ with
$q$ is very slow, and arises exclusively from the lipid mixing
term. In regime (ii) where $1<q<4/3$, the decrease in $\Delta F$ is
faster due to the additional contribution of counterion release.
Finally, in regime (iii) where $q>4/3$, lipid mixing becomes again a
dominant factor, though there is a fixed gain of entropy due to
counterion release.

The key to understand the trends in
fig.~\ref{fig:fig2_iso_entropy}\subfig{a} is to correctly identify the
transition points between the three different regimes.  The transition
from (i) to (ii) occurs at $q=1$, which is the isoelectric point of
the lipoplex, namely the point where the total cationic charge of the
lipids \emph{exactly}\/ matches the negative one of the DNA array: $2
\Phi_+ = \phi_D$.  Therefore, in regime (i) ($q<1$), the internal
solution surrounding the DNA array includes cations [see
eq.~\eqref{eq:sigma_b3}]. Similarly, the external solution facing the
cytoplasm, and the intermediate solutions between the plasma membrane
and the lipoplex, also include cations only [eqs.~\eqref{eq:sigma_b1}
and \eqref{eq:sigma_b2}]. Since the system contains no anions, it is
impossible to release cation-anion pairs, which explains why, in this
regime, the only contribution to the free energy comes from mixing of
the lipids. Equilibrium is achieved when the lipids are evenly
distributed between the four monolayers. In contrast to regime (i), in
regime (ii) ($1<q<4/3$) both the intermediate and the internal
solutions include anions at the initial conditions, while the external
solutions contains cations. Therefore, the decrease in free energy now
involves contributions of both lipid mixing and counterion release.
Detailed calculation shows that in regime (ii), equilibrium is reached
when all the anions are released, while the excess cations accumulate
at the internal water layer around the DNA molecules. Moreover, to
satisfy the conditions of eqs.~\eqref{eq:sigma_b1} and
\eqref{eq:sigma_b2}, the net charge density $\sigma_i$ in monolayers
$i=1,2,3$ must vanish, which means that the mole fractions of CLs and
ALs in each of these layers are the same.  The composition of layer
$i=4$ is different, which implies that lipid mixing is not optimized
in regime (ii). Regime (ii) ends at $q=4/3$, which is the point where
the total charge of the system (including the ALs of the plasma
membrane, the CLs of the lipoplex, and the DNA array) vanishes; i.e.,
when
\begin{equation}
    3 \Phi_+  = \Phi_D + 2\Phi_-.
    \label{eq:phi_+}
\end{equation}
Therefore, at this point, the total numbers of bound cations and
anions is also the same. Further increasing $q$, by increasing the
fraction of the CLs and the number of associated bound anions, we
enter into regime (iii). In this regime, the total gain of free energy
due to counterion release saturates, since it is capped by the number
of cations originally bound to the plasma membrane. The free energy
$\Delta F$ continues to decrease with $q$ since the lipids can now
better mix and attain a more even distribution between monolayers
$i=1-4$. The equilibrium distribution of lipids between the four
monolayers are depicted in
fig.~\ref{fig:fig2_iso_entropy}\subfig{b}. Notice that the composition
of lipids in monolayers $i=2,3$ is always the same, which is
anticipated since any exchange of lipids between these two monolayers
will not influence the charge balance condition of
eq.~\eqref{eq:sigma_b2}.

\begin{figure}
    \centering
    \includegraphics[width=0.4\textwidth]{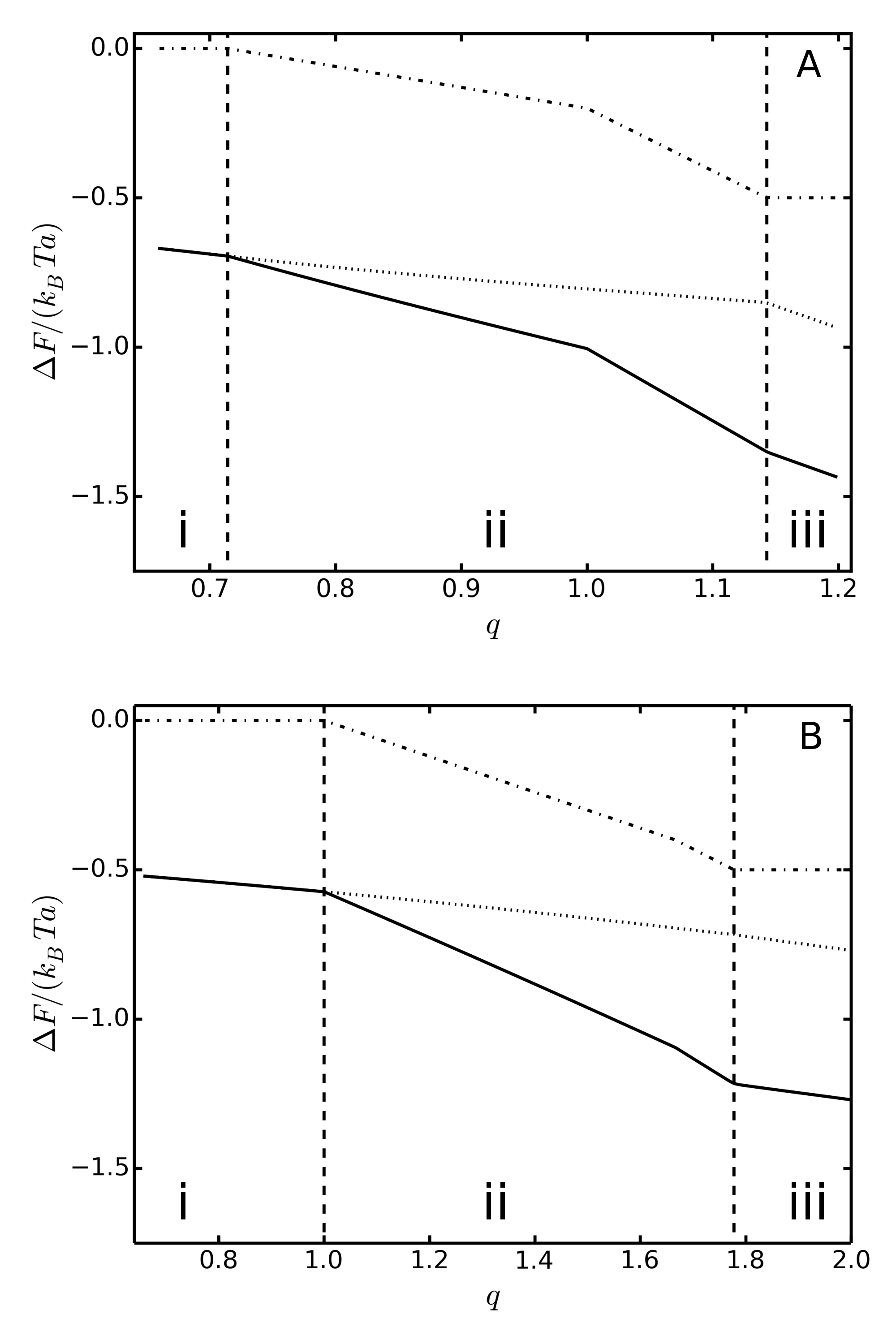}
    \caption{{\bf (A)} The free energy $\Delta F$ (solid line) and the
    partial contributions to $\Delta F$ originating from counterions
    release (dot dashed line) and lipid mixing (dotted line).  Results
    are for a lipoplex with densely packed DNA molecules ($\Phi_D
    =1.4$).  The vertical dashed lines marks the transition points
    between the regimes discussed in the text. {\bf (B)} Same as in {\bf
    (A)} for a lipoplex with loosely packed DNA molecules ($\phi_D =
    0.6$).}  %
    \label{fig:fig3_energy_0307}
\end{figure}

Figure \ref{fig:fig3_energy_0307}\subfig{a} depicts our results for
$\Delta F$ for a lipoplex with more densely packed DNA rods ($\Phi_D =
1.4$). The charge density of the plasma membrane is the same as in
fig.~\ref{fig:fig2_iso_entropy}\subfig{a}, $\Phi_- = 0.5$. The
characteristics of fig.~\ref{fig:fig3_energy_0307}\subfig{a} are very
similar to the those observed in
fig.~\ref{fig:fig2_iso_entropy}\subfig{a}. One noticeable difference
is that regime (ii) starts below the isoelectric point $q<1$, at
$q=\Phi_D^{-1} \simeq 0.71$. As in the previously discussed case, in
regime (i) the initial state of the system includes only cations. In
regime (ii) the intermediate water layer contains anions, which are
released upon reaching equilibrium. The kink at the isoelectric point
is due to the fact that for $q>1$, the internal solution also contains
anions.  The transition between regions (ii) and (iii) is at $q\simeq
1.14$, as dictated by eq.~\eqref{eq:phi_+}. In regime (iii), the
contribution of counterions release to $\Delta F$ is fixed by the
amount of cations present in the system.

Figure \ref{fig:fig3_energy_0307}\subfig{b} depicts our results for
$\Delta F$ for a lipoplex with more loosely packed DNA rods ($\Phi_D =
0.6$), with a plasma membrane of charge density $\Phi_- = 0.5$. Here,
the transition from (i) to (ii) is at the isoelectric point $q=1$
which, as noted above, is where anions first appear at the internal
layer next to the DNA. The kink happens at $q=\Phi_D^{-1} \simeq1.67$
above which, the intermediate water layer contains anions at the
initial state. The transition from (ii) to (iii)
occurs at $q\simeq1.78$, as predicted by eq.~\eqref{eq:phi_+}.

The free energy calculations reported in
figs.~\ref{fig:fig2_iso_entropy} and \ref{fig:fig3_energy_0307}
demonstrate the inherent instability of the entrapped lipoplex,
triggered by its interactions with the enveloping plasma membrane. The
latter constitutes a reservoir of ALs that can mix with the CLs of the
lipoplex. Lipid mixing occurs through ``flip-flop'' events which, in
general, are slow especially when lipids transfer between distinct
bilayers (as opposed to lipids moving between monolayers of the same
membrane, which is probably somewhat faster). The exchange of lipids
between the plasma and lipoplex membranes may cause these two
membranes to fuse \cite{Ewert:2004} - a scenario that we have thus far
not taken into account.  Fusion is thermodynamically favorable since
it reduces the number of participating monolayers from $i=4$ to $i=2$
and, thus, it further increases the lipid mixing entropy. However, it
comes with the cost of bending energy. Crossing the associated energy
barrier is what primarily determines the rate of successful endosomal
escape and sets the TE (transfection efficiency). Experimentally, it
is known that the TE of lamellar complexes grows exponentially with
the cationic charge density of the complex, $\Phi_+ = (q
\Phi_D)/2$. This observation supports the picture of activated fusion
where ${\rm TE} \sim \exp \left (-\Delta F/k_BT \right)$, and
\begin{equation}
    \Delta F = a \kappa  - b \Phi_+ +c,
    \label{eq:Safinya}
\end{equation}
where $\kappa$ is the bending rigidity of the bilayers, while $a$,
$b$, and $c$ are parameters, the value of which may depend on the
molecular conditions inside the endosome (see eq.~(2) in
\cite{Ewert:2004}). The first term in eq.~\eqref{eq:Safinya}
represents the curvature energy cost of the fusion which, to a good
approximation, is independent of the charge densities. The second term
has been previously attributed to the electrostatic attraction between
the plasma membrane and the complex. The last term accounts for other
effects, e.g., the capacity of the low-pH environment of the endosome
to disrupt the lipid bilayer. Our study reveals that the origin of
second term is actually {\em not}\/ energetic but entropic. The free
energy gain $\Delta F$ due to lipid mixing and the associated
counterion release at the second stage of the transfection process
(see solid curves in fig.~\ref{fig:fig2_iso_entropy} and
\ref{fig:fig3_energy_0307}) grows piecewise linearly with
$\Phi_+$. This linear dependence is simply a reflection of the fact
that when the lipoplex contains a higher fraction of CLs, the
potential entropic gain involved in ideal mixing of lipids and
counterions release is larger.

Once fusion occurs, a hole opens that connects the cytoplasm and the
internal water layer containing the first DNA array of the
lipoplex. This allows for influx of positively charged
(macro)molecules, e.g., unstructured peptides, that are able to
condense the DNA molecules and release them to the cytoplasm
\cite{footnote3}. Removing the first DNA layer leaves us with a
smaller lipoplex whose composition is similar to the original
one. Interactions of this positively charged complex with negatively
charged components of the cell may cause renewed thermodynamic
instability and lead to further degradation of the CL-DNA complex.

To conclude, we use a simplified model to study the thermodynamics of
transfection by CL-DNA complexes (lipoplexes). The formation of these
complexes is known to be driven by the increase in the translational
entropy of the counterions that are released to the bulk solution when
the oppositely charged membranes and DNA molecules associate
together. The same counterion release mechanism mediates (at least
partially) the adhesion of the lipoplex to the cell plasma membrane,
which initiates the transfection process. In this work, we argue that
the contact between the lipoplex external bilayer and the plasma
membrane triggers thermodynamic instability that leads to lipoplex
degradation, which is essential for the transfection process to
proceed.  The thermodynamic instability of the entrapped lipoplex is
of entropic origin: It stems from the fact that the lipid composition
of the lipoplex and the plasma membrane is different and, therefore,
mixing of these lipids increases the configurational entropy of the
system. Since the two membranes are oppositely charged, the mixing of
lipids has another effect - it reduces the charge density of the
membranes. This enables further counterion release and a further
decrease in the free energy. Thus, the counterion release mechanism
which has been identified as the thermodynamic driving force for
formation of various supramolecular structures \cite{harriers-review},
is here used to explain the disassembly of such structures.

Despite the gross simplicity of our model and the fact that it ignores
specific molecular details, it successfully predicts a roughly {\em
linear}\/ increase in the free energy gain with the mole fraction of
CLs in the complex, which explains the observed {\em exponential}\/
increase in transfection efficiency of lamellar complexes with the
charge density. The model is based on a mean-field picture and
replaces the lipid monolayers with uniformly charged flat
surfaces. This modeling approach is routinely used in theoretical
studies of electrostatic effects in soft matter systems.  We avoid
solving the Poisson-Boltzmann (PB) equation explicitly, and instead
associate each released counterion with a free energy gain of $1
k_BT$. By solving the PB equation, a more accurate value may be
obtained (which may depend on the water layer from where the
counterion is released), but the result is only expected to be
different by a factor of order unity.  What might be the boldest
approximation in our model is the replacement of the DNA array with a
uniformly charged surface as well. By employing this picture, we
essentially ignore two entropic contributions of opposite sign: (i)
The CLs in the monolayers facing the DNA arrays are expected to
accumulate near the DNA rods, which lowers their mixing entropy. (ii)
The space available to the ions surrounding the DNA molecules is quite
small, which implies that the entropic gain involving in their release
may be higher than assumed by the model. The order of magnitude of
these effects is comparable to the other contributions discussed
here. Therefore, although we do not expect these two entropic
contributions to cancel each other, we also do not expect them to
dominate and modify the picture presented here.

This work was supported by the Israel Science Foundation (ISF) through grant
no.~1087/13.


\begin{thebibliography}{99}
\bibitem{footnote1} In some cases (see text), these constaints may
    bring anoins to the internal solution of the DNA array.
\bibitem{footnote2} We assume no mixing with the lipids of monolayer
$i=6$, which is separated from the two external bilayers (monolayers
$i=1-4$) by the DNA array and the surrounding internal water layer.
\bibitem{footnote3} Upon attachment to the DNA, the counterions
surrounding the charged macromolecules can be released, which provides
an even greater driving force to this process.
\end{thebibliography}
\end{document}